\documentclass{moriond}

\def\be{\begin{equation}}
\def\ee{\end{equation}}
\def\bea{\begin{eqnarray}}
\def\eea{\end{eqnarray}}

\usepackage{xspace}
\newcommand{\NNLOJET}{NNLO\protect\scalebox{0.8}{JET}\xspace}

\newcommand{\Photo}{}

\begin{document}

\title{Differential single jet inclusive production at Next-to-Next-to-Leading Order in QCD}

\author{J.Currie$^{1}$, A.Gehrmann-De Ridder$^{2}$, T. Gehrmann$^{3}$, E.W.N Glover$^{1}$, A. Huss$^{2}$, J. Pires$^{4}$~\footnote{Speaker.}}

\address{
$^1$ Institute for Particle Physics Phenomenology, University of Durham, Durham, DH1 3LE, UK\\
$^2$ Institute for Theoretical Physics, ETH, CH-8093 Z\"urich, Switzerland \\
$^3$ Department of Physics, Universit\"at Z\"urich, Winterthurerstrasse 190, CH-8057 Z\"urich, Switzerland \\
$^4$ Max-Planck-Institut f\"ur Physik, F\"ohringer Ring 6 D-80805 Munich, Germany
}

\maketitle\abstracts{
In this talk we present the recent calculation in all partonic channels 
of the fully differential single jet inclusive cross section at Next-to-Next-to-Leading Order in QCD.
We discuss the size and shape of the perturbative corrections as a function of the functional form of the
renormalisation and factorisation scales and compare the predictions at NLO and NNLO
to the available ATLAS 7 TeV data. We find significant effects at low-$p_T$ due to changes in the
functional form of the scale choice whereas at high-$p_T$ the two most common scale choices 
in the literature give identical results and the perturbative corrections lead to a substantial
reduction in the scale dependence of the theoretical prediction at NNLO. 
}

\section{Introduction}
Single jet inclusive and dijet observables are the most fundamental QCD processes measured
at hadron colliders. They probe the basic parton-parton scattering in $2\to2$ kinematics, and
thus allow for a determination of the parton distribution functions (PDFs) in the proton and for a direct
probe of the strong coupling constant $\alpha_s$ up to the highest new energy scales that can be attained
in collider experiments.

In the single jet inclusive cross section each identified jet in an event contributes individually to the cross section.
In particular, all subleading jets that pass the jet fiducial cuts in the same event are booked
at the histogram level together with the leading jet. This cross section has been studied as function of the transverse
momentum $p_T$ and absolute rapidity $|y|$ of the jets and precision measurements
performed recently by the ATLAS and CMS collaborations at $\sqrt{s}=8$ TeV and $\sqrt{s}=13$ TeV have been presented at this conference~\cite{Zuzana}. 
The discriminating power of the hadron collider jet data to constrain the gluon and valence quark PDFs
has been demonstrated in Ref.~\cite{CMS8TeV} where uncertainties on the gluon PDF at high-$x$ 
(relevant to increase the precision for Higgs, top and other SM measurements and BSM searches) 
are significantly reduced once the CMS $\sqrt{s}=8$ TeV jet measurements are included in the determination of the PDFs. 

Theoretical predictions for this observable accurate to NLO
in QCD have been obtained in~\cite{jetsNLO} (with the inclusion of shower effects in~\cite{jetsNLOPS}) while NLO corrections in the
electroweak theory were obtained in~\cite{jetsEW}.
The ATLAS and CMS analysis of jet data show that QCD is a well established theory however,
many regions of phase space probed by the kinematics of jet production are not well described by current theory predictions.
In particular the level of agreement with NLO theory varies between the different PDF sets and for many experimentally accessible
cross section bins is limited by large scale uncertainties in the NLO theory prediction. The 
latest theoretical development in the description of jet production at a hadron collider is the 
calculation of the NNLO QCD corrections to the single jet inclusive cross section~\cite{jetsNNLO}. In this talk
we present these very recent results as a new approach to look at jet data at the LHC.

\section{Results at NNLO}
\begin{figure}
\begin{minipage}{0.5\linewidth}
\centerline{\includegraphics[width=1.0\linewidth]{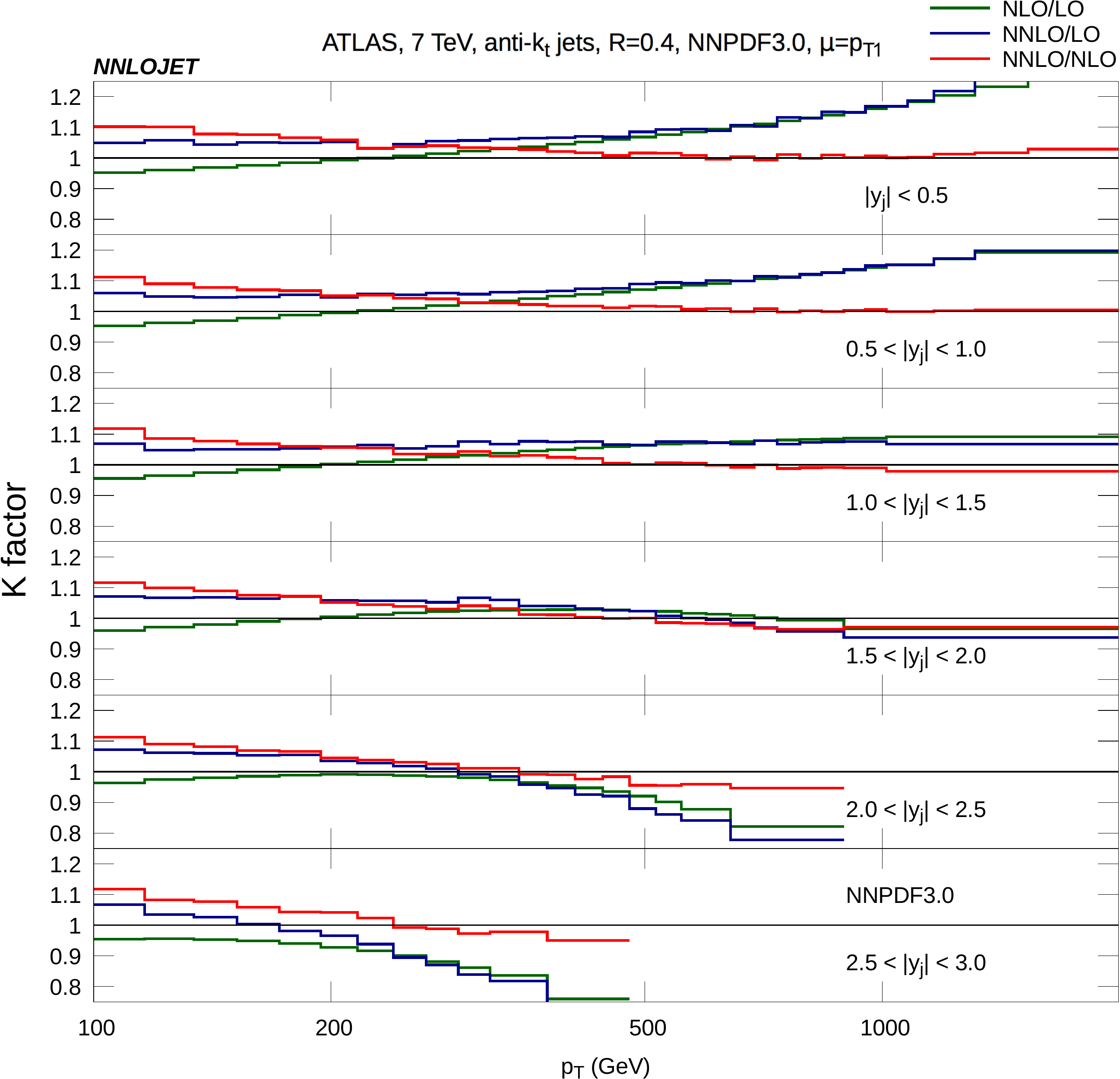}}
\end{minipage}
\hfill
\begin{minipage}{0.5\linewidth}
\centerline{\includegraphics[width=1.0\linewidth]{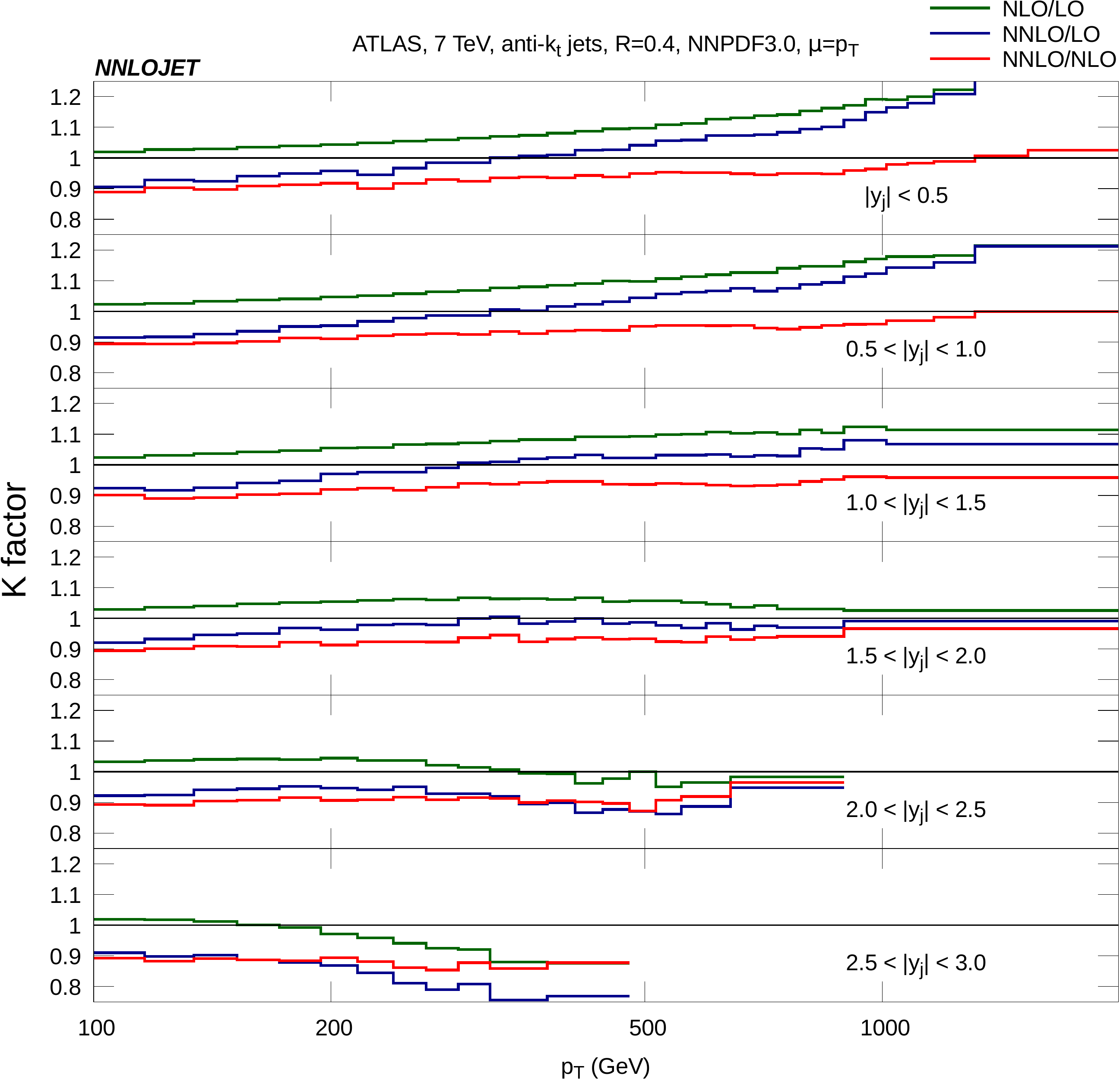}}
\end{minipage}
\hfill
\caption[]{NLO/LO (green), NNLO/NLO (red) and NNLO/LO (blue) $k$-factors for jet production at $\sqrt{s}=7$ TeV. The lines correspond to the double differential
   $k$-factors (ratios of perturbative predictions
  in the perturbative expansion) for $p_T > 100$~GeV and across six rapidity $|y|$ slices. Lines correspond to theoretical predictions evaluated with NNLO PDFs
  from NNPDF3.0 and central scale choice $\mu=p_{T1}$ (left plot) and $\mu=p_{T}$ (right plot).}
\label{fig:KF}
\end{figure}

To perform our calculation we have employed the antenna subtraction scheme~\cite{ant} for the analytic
cancellation of infra-red (IR) singularities at NNLO. As demonstrated in~\cite{RR,RV,VV}, using the antenna subtraction
scheme the explicit $\epsilon$-poles in the dimension regularization parameter of one- and two-loop matrix
elements are cancelled in analytic and local form against the $\epsilon$-poles of
the integrated antenna subtraction terms. All predictions presented in this talk have been obtained 
with the parton level generator \NNLOJET which implements the antenna subtraction scheme to compute 
fully differential jet cross sections at NNLO in QCD.

The results presented here are for the experimental setup ($p_T$ and rapidity bin widths) used by the ATLAS~\cite{ATLAS} collaboration
for the $\sqrt{s}=7$~TeV 4.5 fb$^{-1}$ data set with jets reconstructed using the anti-$k_{T}$ jet algorithm with $R=0.4$. 
The cuts imposed on the jet data include all jets found with $p_{T}\ge 100$~GeV
and $|y|<3$. The theoretical calculation uses the NNPDF3.0 NNLO PDF set with $\alpha_{s}(M_{Z}^2)=0.118$ for LO, NLO
and NNLO contributions. Similarly to the analysis performed by ATLAS~\cite{ATLAS} 
we set the renormalisation scale, $\mu_R$, and the factorisation scale, $\mu_F$, in the theory prediction equal
to the leading jet transverse momentum $p_{T1}$ for each event. Additionally we present results using
the individual jet transverse momentum $p_T$ at the event level as the $\mu_R$ and $\mu_F$ scales for each jet's 
contribution to the single jet inclusive cross section. For the leading jet in the event this scale
is identical to $p_{T1}$ and so its contribution is insensitive to the scale choice between $p_T$ and $p_{T1}$. 
Similarly, 2-jet events where the jets are balanced in $p_T$ cannot generate any difference
as $p_T$ = $p_{T1}$ = $p_{T2}$. Away from these jet configurations, the subleading jets
will have smaller $p_T$ than the leading jet in the event and so choosing the
individual jet $p_T$ as the theoretical scale will mean that the scale used to
calculate the weight associated with a jet will on average be smaller than
the scale $p_{T1}$. 

For these reasons at the LO the two scale choices generate the same prediction and similarly, for all events at higher order
that have LO kinematics there is no difference between the two scale choices. In particular at high-$p_T$
the scale choices once again converge as is to be expected for the largely back-to back
configurations found at high-$p_T$. Kinematical configurations where the scale choices do not coincide
are events with three or more hard jets and events with hard emissions outside the
jet fiducial cuts that generate an imbalance in $p_T$ between the leading and subleading jets in the event. 

\begin{figure}
\begin{minipage}{0.5\linewidth}
\centerline{\includegraphics[width=1.0\linewidth]{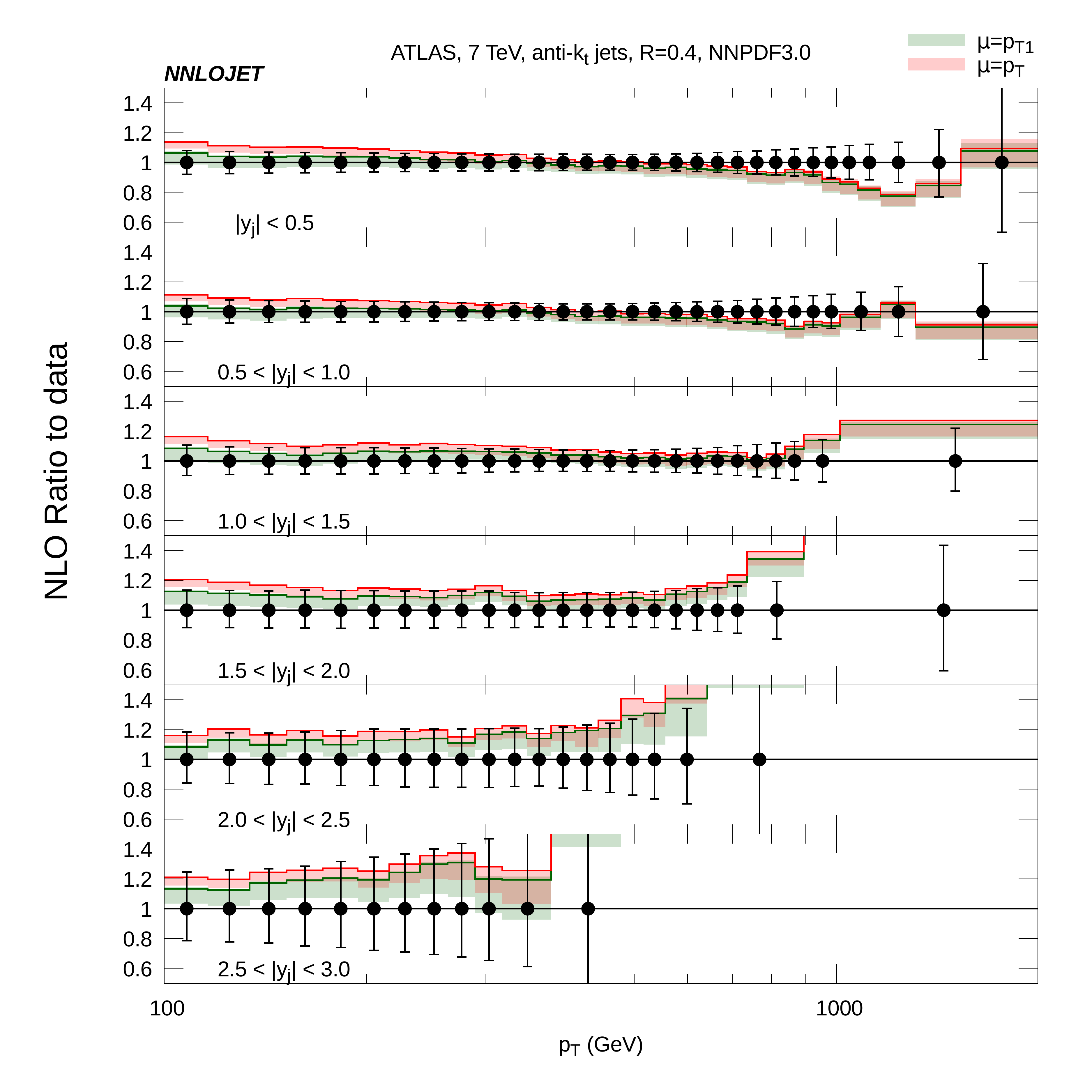}}
\end{minipage}
\hfill
\begin{minipage}{0.5\linewidth}
\centerline{\includegraphics[width=1.0\linewidth]{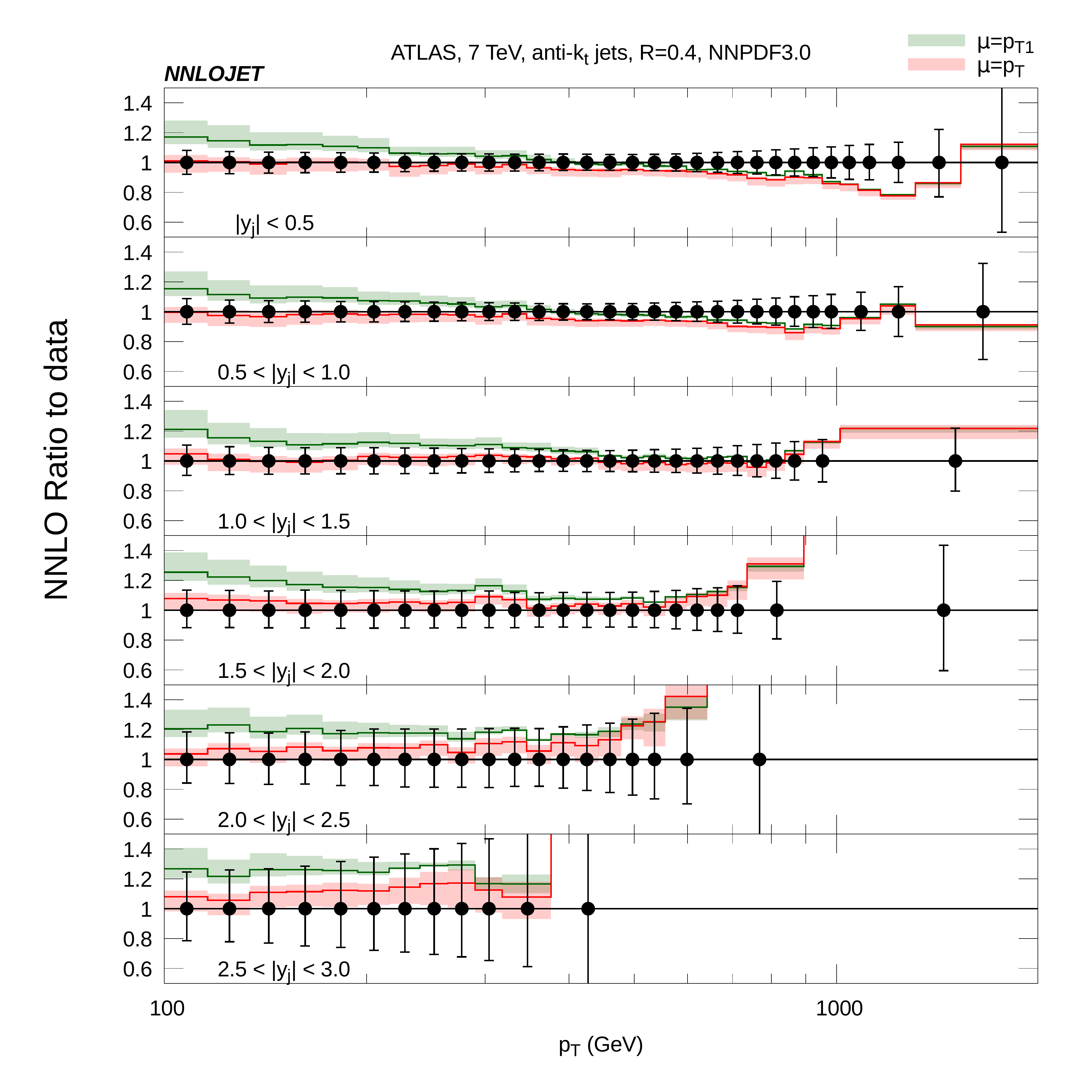}}
\end{minipage}
\hfill
\caption[]{NLO predictions (left plot) and NNLO predictions (right plot) normalized to data 
for two different scale choices, individual jet $p_T$ (red) and leading jet $p_T$ (green). The bands correspond to the
variation of $\mu=\mu_R=\mu_F$ by factors of 0.5 and 2 about the central scale choice.}
\label{fig:ratiotodata}
\end{figure}

In Fig.~\ref{fig:KF} we show the potential for the NNLO correction to change
the shape of the distribution relative to NLO. As a function of $p_{T}$ in six rapidity slices
we show the $k$-factors for NLO/LO, NNLO/NLO and NNLO/LO for a perturbative expansion using the scale 
$\mu=p_{T1}$ (left plot) and $\mu=p_{T}$ (right plot). Using the $p_{T1}$ scale choice we observe that the NNLO
coefficient increases the NLO result at the 10\% level at low $p_{T}$ for all rapidity slices while the  
effects at high-$p_{T}$ are small. The shape of the NNLO/NLO $k$-factor is getting steeper in the
forward rapidity slices. On the other hand using the $p_{T}$ scale choice we see that at low-$p_{T}$ the
NNLO/NLO $k$-factor provides a negative 10\% correction, decreases in magnitude at higher $p_T$ and the shape
of the NNLO/NLO $k$-factor flattens in the forward rapidity slices. The difference in the shape
of the $k$-factor between the two scale choices seems to indicate that there is a potential interplay 
between the scale choice in the theory prediction and a consistent fit of jet data in PDF's for all 
rapidity slices simultaneously. For this reason a detailed study of the effects of the single
jet inclusive datasets and NNLO theory predictions on PDF fits is required for more substantive conclusions

In Fig.~\ref{fig:ratiotodata} we present the comparison of the theory predictions at NLO and NNLO with the ATLAS data
for the two scale choices. Looking at the results at NLO on the left side of the figure, we find small differences in the central value
of the predictions at low-$p_T$ which are inside the scale dependence of the NLO prediction, estimated by varying
both central scale choices by a factor of two and one half and represented by the thickness of the
bands. We observe that both scale choices show an asymmetric scale band where the central value of the prediction sits at the top of the band.
Moreover the scale uncertainty of the NLO prediction at low-$p_T$ is underestimated due to the turnover of the NLO coefficient
from negative to positive. Scale uncertainties at high-$p_T$ are around~20\% rising to~40\% for forward jets. When comparing
the results with the data we do not include non-perturbative effects; they are quantified
in Ref.~\cite{ATLAS} and found to be a 2\% effect in the lowest $p_T$ bin and at most a 1\% effect in all other bins.

In the same figure on the right side we compare the data with the predictions at NNLO in QCD. In comparison with the results
at NLO we observe that both scale choices show a more reliable symmetric scale variation. The scale uncertainty at NNLO
is at the~10\% level at low-$p_{T}$ and at the percent level at high-$p_{T}$. At high-$p_{T}$ the predictions with $\mu=p_{T1}$ and 
$\mu=p_{T}$ coincide whereas at the low-$p_{T}$ we observe significant differences which are outside the NNLO scale
variation band. At low-$p_T$ we find the behaviour somewhat different to NLO: the NNLO correction for the $p_{T1}$
scale moves the prediction away from the data, with which it was consistent at NLO; whereas using the $p_T$ scale brings the
NNLO prediction in line with the data with which there was some tension at NLO. The significant effect of this 
scale ambiguity on the NNLO predictions, and the lack of a theoretically well motivated preference motivates 
further study of this issue.

\section{Conclusions and outlook}
In this talk we reported the first results on the NNLO QCD radiative corrections 
to the single jet inclusive cross section at hadron colliders. In the single jet inclusive cross section
all jets in the event that pass the jet fiducial cuts contribute to the jet transverse momentum
distribution. For this reason two scale choices for the theoretical predictions in perturbative QCD 
have been considered in the literature; the leading transverse 
momentum scale choice $\mu=p_{T1}$ for all jets in the event or the individual jet
transverse momentum $\mu=p_{T}$ of each jet in the event. In the medium to high-$p_T$ jet
transverse momentum distribution the two scale choices yield identical results and the scale
uncertainties are at the percent level at NNLO, a substantial reduction with respect to NLO. 
At low-$p_{T}$ we observe significant differences in the NNLO prediction outside 
the NNLO scale variation band. This motivates further studies using LHC jet data 
to understand jet production at hadron colliders, including studies of scale choice, 
jet shape, cone size and different PDF sets. In particular
it would be desirable to have a consistent description of jet data at NNLO for all 
jet datasets at low and high-$p_{T}$ in the central and forward rapidity regions for multiple
jet $R$ cone sizes. 

\section*{Acknowledgments}
The authors thank Xuan Chen, Juan Cruz-Martinez,
Tom Morgan and Jan Niehues for useful discussions and their many contributions to the \NNLOJET code.
We gratefully acknowledge the assistance provided by Jeppe Andersen utilizing the computing resources 
provided by the WLCG through the GridPP Collaboration. 
This research was supported in part by the UK Science and Technology Facilities Council, in part by the
Swiss National Science Foundation (SNF) under contracts 200020-162487 and CRSII2-160814, in part by the Research 
Executive Agency (REA) of the European Union under the Grant Agreement PITN-GA-2012-316704 (``HiggsTools'')
and the ERC Advanced Grant MC@NNLO (340983).

\end{document}